\documentclass[conference]{IEEEtran}
\usepackage{filecontents}
\usepackage{comment}
\usepackage{booktabs}
\usepackage{algorithm}
\usepackage{algorithmic}
\usepackage{multirow}
\usepackage{listings}
\usepackage{xcolor}
\usepackage{url}
\usepackage{graphicx}
\ifCLASSINFOpdf
\else
\fi
\definecolor{mGreen}{rgb}{0,0.6,0}
\definecolor{mGray}{rgb}{0.5,0.5,0.5}
\definecolor{mPurple}{rgb}{0.58,0,0.82}

\lstdefinestyle{CStyle}{
    backgroundcolor=\color{white},   
    commentstyle=\color{mGreen},
    keywordstyle=\color{blue},
    numberstyle=\tiny\color{mGray},
    stringstyle=\color{mPurple},
    basicstyle=\footnotesize,
    frame=single,
    breakatwhitespace=false,         
    breaklines=true,                 
    captionpos=b,                    
    keepspaces=true,                 
    numbers=left,                    
    numbersep=5pt,                  
    showspaces=false,                
    showstringspaces=false,
    showtabs=false,                  
    tabsize=2,
    language=C
}

\begin{document}
%

\title{Unintentional Security Flaws in Code: Automated Defense via Root Cause Analysis}


\author{
\IEEEauthorblockN{Nafis Tanveer Islam}
\IEEEauthorblockA{University of Texas at San Antonio\\
nafistanveer.islam@utsa.edu}
\and
\IEEEauthorblockN{Mazal Bethany}
\IEEEauthorblockA{University of Texas at San Antonio\\
mazal.bethany@utsa.edu}
\and
\IEEEauthorblockN{Dylan Manuel}
\IEEEauthorblockA{University of Texas at San Antonio\\
dylan.manuel@utsa.edu}
\and
\IEEEauthorblockN{Murtuza Jadliwala}
\IEEEauthorblockA{University of Texas at San Antonio\\
murtuza.jadliwala@utsa.edu}
\and
\IEEEauthorblockN{Peyman Najafirad}
\IEEEauthorblockA{University of Texas at San Antonio\\
peyman.najafirad@utsa.edu}
}


%


\IEEEoverridecommandlockouts
\makeatletter\def\@IEEEpubidpullup{6.5\baselineskip}\makeatother
\IEEEpubid{\parbox{\columnwidth}{
    Network and Distributed System Security (NDSS) Symposium 2024\\
    26 February - 1 March 2024, San Diego, CA, USA\\
    ISBN 1-891562-93-2\\
    https://dx.doi.org/10.14722/ndss.2024.23xxx\\
    www.ndss-symposium.org
}
\hspace{\columnsep}\makebox[\columnwidth]{}}

\maketitle

\begin{abstract}
Software security remains a critical concern, particularly as junior developers, often lacking comprehensive knowledge of security practices, contribute to codebases. This situation underscores the need for tools that empower these developers to identify and address vulnerabilities proactively. While there are tools to help developers proactively write secure code, their actual effectiveness in helping developers fix their vulnerable code remains largely unmeasured. Moreover, these approaches typically focus on classifying and localizing vulnerabilities without highlighting the specific code segments that are the root cause of the issues, a crucial aspect for developers seeking to fix their vulnerable code.
To address these challenges, we conducted a comprehensive study evaluating the efficacy of existing methods in helping junior developers secure their code. Our findings across five types of security vulnerabilities revealed that current tools enabled developers to secure only 36.2\% of vulnerable code. Questionnaire results from these participants further indicated that not knowing the code that was the root cause of the vulnerability was one of their primary challenges in repairing the vulnerable code.
Informed by these insights, we developed an automated vulnerability root cause (RC) toolkit called T5-RCGCN, that combines T5 language model embeddings with a graph convolutional network (GCN) for vulnerability classification and localization. Additionally, we integrated DeepLiftSHAP to identify the code segments that were the root cause of the vulnerability.
We conduct our study on a total of 56 junior developers and three source code vulnerability datasets, and demonstrated that T5-RCGCN improved developers' ability to proactively secure code by 28.9\% compared to previous methods. Notably, we also observed educational benefits, as developers who had used our tool showed enhanced capability in securing code without assistance, having gained a deeper understanding of vulnerability root causes. We measured a 17.0\% improvement in this learning outcome compared to other source code vulnerability analysis tools. These results show the potential of our approach in both immediate code security improvement and long-term developer skill enhancement.
\end{abstract}


%

\section{Introduction}
\label{1_introduction}
Open-source software (OSS) has become a cornerstone of modern technology, powering systems from IoT platforms \cite{al2022idetect} to critical software supply chains \cite{ladisa2023sok}. However, this widespread adoption has also made OSS a prime target for cyber adversaries. High-profile breaches, such as the compromise of SolarWind's Orion platform \cite{peisert2021perspectives}, which affected approximately 18,000 stakeholders including government bodies and critical infrastructure providers, underscore the severe consequences of software vulnerabilities. These incidents highlight the critical nature of software security, particularly in widely-used libraries \cite{log4j}, where vulnerabilities can lead to extensive service disruptions \cite{synopsysrisc, log4j}.
The rapid evolution of technology, including the emergence of AI-powered coding assistants like GPT-4 \cite{chatgpt} and GitHub Copilot \cite{githubcopilot}, has further complicated the software security landscape. While these tools can accelerate development, they also introduce new security challenges \cite{khoury2023secure, sandoval2022security}, often generating code that may contain vulnerabilities. This is particularly concerning when such code is utilized by developers with limited security knowledge.
Current statistics paint a worrying picture, with nearly 40\% of open-source code potentially falling short of stringent security standards \cite{pearce2022asleep}. This situation is exacerbated by the fact that many developers, especially those in junior positions, often lack comprehensive knowledge of security practices. Recent research has highlighted significant gaps in developers' understanding of security practices and limited access to security expertise \cite{green2016developers, acar2016you, assal2019think, weir2020needs}.
These factors collectively underscore the critical need for tools and approaches that can empower developers, particularly junior developers, to identify and address vulnerabilities proactively.

Recent years have seen significant innovations in software security tools aimed at assisting developers. These advancements include automated vulnerability detection techniques using transformer-based models \cite{nguyen2021regvd, islam2023unbiased}, graph convolutional networks \cite{guo2020graphcodebert}, as well as automated vulnerability repair (AVR) systems \cite{pearce2023examining, joshi2023repair, chen2022neural} powered by generative models \cite{2020t5, wang2021codet5}. These tools offer promising capabilities in identifying and potentially fixing vulnerabilities in static source code.
However, a critical gap remains in our understanding of these tools' practical impact: their actual effectiveness in helping developers to proactively secure their code remains largely unmeasured. This lack of comprehensive evaluation leaves us without clear insights into how well these advanced techniques translate into tangible improvements in code security when applied by developers in real-world scenarios.
Moreover, these tools focus on vulnerability classification and localization. This leaves a critical gap: the identification of the code that is the root cause of the vulnerability. This limitation is particularly problematic for junior developers who may lack the expertise to interpret and act on vulnerability classification and localization alone \cite{acar2016you, green2016developers}. The challenge is further compounded in large-scale applications with millions of lines of code, where identifying the precise origin of a vulnerability becomes increasingly complex \cite{bellare1997forward, capobianco2019employing, han2020unicorn}. These two significant gaps - the lack of measured effectiveness of tools to help developers secure their code and the absence of root cause identification - call for focused research in the field of software security. Specifically, this research should aim to: (1) quantitatively evaluate how effectively existing tools help developers in securing their code; and (2) develop tools that not only identify and localize vulnerabilities but also pinpoint the specific code segments that are the root cause of these issues.

To address these challenges, we conducted a comprehensive study evaluating the efficacy of existing methods in helping junior developers secure their code. Our investigation focused on five types of common security vulnerabilities, revealing that current tools enabled junior developers to secure only 36.2\% of vulnerable code. This finding underscores the significant room for improvement in developer-assistance tools. Furthermore, through questionnaires administered to study participants, we identified a critical gap: developers frequently cited not knowing the root cause of vulnerabilities as a primary obstacle in repairing vulnerable code.

Informed by these findings, we developed T5-RCGCN, a novel tool designed to help developers understand, analyze, and fix source code vulnerabilities. T5-RCGCN combines the power of T5 language model embeddings with a graph convolutional network (GCN) for enhanced vulnerability classification and localization. This approach leverages the contextual understanding of large language models and the structural analysis capabilities of graph-based methods. Crucially, we integrated DeepLiftSHAP into our tool, enabling the identification of specific code segments that serve as the root cause of vulnerabilities. This feature directly addresses the key challenge reported by developers in our initial study, providing them with better context for effective code repair. To evaluate T5-RCGCN's effectiveness, we conducted a series of comprehensive experiments. First, we assessed T5-RCGCN's performance on vulnerability classification and localization using two common source code vulnerability datasets. The results demonstrated state-of-the-art performance, with T5-RCGCN achieving an average vulnerability classification F1 score 11.2\% higher than the next best-performing methods. For vulnerability localization, T5-RCGCN outperformed existing top methods by 16.5\% on the Intersection over Union (IoU) metric. To further demonstrate the robustness and generalizability of T5-RCGCN to real-world vulnerabilities, we gather six open-source IoT repositories and detect 24 n-day vulnerabilities, which was 10 more than the next best performing tool.
We then investigated how the integration of DeepLiftSHAP for root cause identification impacted junior developers' ability to secure vulnerable code. This experiment revealed a significant improvement, with T5-RCGCN enabling a 24\% increase in junior developers' capacity to write secure code compared to other source code analysis tools.
Lastly, we examined the educational benefits of T5-RCGCN by comparing how much junior developers improved their understanding of secure coding practices after using our tool versus other source code analysis tools. The results were promising, showing that developers who used T5-RCGCN demonstrated a 9\% greater improvement in their ability to secure code without assistance compared to those who had used other tools. This finding suggests that T5-RCGCN not only aids in immediate code security improvement but also contributes to long-term developer skill enhancement by providing a deeper understanding of vulnerability root causes.

Our work makes the following contributions:
\begin{itemize}

    \item We conducted a novel study evaluating the effectiveness of software security tools in assisting junior developers across five types of vulnerabilities, revealing the limited ability of existing tools to help junior developers fix code vulnerabilities.

    \item We introduce T5-RCGCN, a novel tool that combines T5 language model embeddings with a graph convolutional network for enhanced vulnerability classification and localization. T5-RCGCN integrates DeepLiftSHAP to identify specific code segments that are the root cause of vulnerabilities, addressing a critical gap in existing tools.

    \item We performed a follow-up assessment to measure the educational impact of software security tools on junior developers to write secure code without the assistance of tools, and find that our proposed system has the potential not only for immediate code security enhancement but also for long-term developer skill improvement.

\end{itemize}

Our data source code and survey materials are made publicly available here. \footnote{\url{https://anonymous.4open.science/r/Threat_Detection_Modeling-BB7B/README.md} }

We structure the paper as follows: In Section \ref{2_background} we detail the background and related work, outlining the gaps in prior works which our research plan to fill. Next, in Section \ref{3_Motivation}, we motivate our research by studying junior developers' code security practices and discovering that existing source code vulnerability analysis tool are limited in helping them secure their code. Following this, we have Section \ref{4_preliminaries} to outline the threat model studied in this research. After that, we present T5-RCGCN, our proposed tool in \ref{5_proposed_system}, whose design is informed by the finding in Section \ref{3_Motivation} that junior developers desire tools that help them discover the root cause of code vulnerabilities. Then, in Section \ref{6_experiments_discussion}, we evaluate the code vulnerability detection and localization abilities of T5-RCGCN against other top performing source code vulnerability analysis tools before demonstrating the benefit seen by junior developers who used T5-RCGCN. We then outline the limitations and future work in Section \ref{8_limitations}, before concluding our research in Section \ref{conclusion}

\section{Background and Related Work}
\label{2_background}


\subsection{Code Vulnerability Classification}
Code vulnerability classification primarily involves determining whether security vulnerabilities exist in source code and, if present, classifying them according to Common Weakness Enumeration (CWE) categories. The field has evolved from conventional machine learning (ML) techniques using manual feature engineering to more flexible deep learning-based solutions with automated feature extraction.
Lin et al. \cite{lin2020software} highlight ML methods as a promising avenue for automated vulnerability discovery. Deep learning-based solutions like VulDeepecker \cite{li2018vuldeepecker} and $\mu$VulDeepecker \cite{zou2019mu} offer universal applicability, though they still rely heavily on feature engineering.
Advancements in vulnerability detection include Devign \cite{zhou2019devign}, VulBERTa \cite{hanif2022vulberta}, and ReVEAL \cite{chakraborty2021deep}, which employ Code Property Graph (CPG) \cite{yamaguchi2014modeling} techniques for improved flexibility. Concurrently, Abstract Syntax Tree (AST) based approaches, proposed by Bilgin et al. \cite{bilgin2020vulnerability} and others \cite{lin2019software,li2021towards,dam2018automatic}, aim to preserve syntactic information during detection.
Transformer-based models, such as RoBERTa \cite{liu2019roberta}, have been adapted for vulnerability detection in source code, as demonstrated by VulBERTa \cite{hanif2022vulberta} and Thapa et al. \cite{thapa2022transformer}. More recently, Islam et al. \cite{islam2023unbiased} proposed a semantic understanding approach to programming languages for vulnerability classification and detection.

\subsection{Code Vulnerability Localization}
Code Vulnerability localization is defined as finding the line from the source code, where the program will break or cause a runtime error. The task of localizing vulnerabilities within source code has seen the application of various methods, from traditional rule-based static analysis tools to innovative deep learning techniques. While tools like Cppcheck \cite{cppcheck}, FlawFinder \cite{flawfinder}, RATS \cite{rats}, and Infer \cite{infer} provide direct approaches to vulnerability localization, their high false positive and false negative rates \cite{yamaguchi2015pattern} underscore the need for more reliable methods. A promising avenue is provided by the use of deep learning, as evidenced by the fine-grained vulnerability detection and locator systems proposed by Vuldeelocator \cite{li2021vuldeelocator} and DeepLineDP \cite{pornprasit2022deeplinedp}, leveraging bidirectional Recurrent Neural Networks (RNN). Additionally, the ensemble graph-transformer learning approach by VELVET \cite{ding2021velvet} and the non-conventional explainability technique by LineVul \cite{fu2022linevul} demonstrate advances in detecting and localizing vulnerabilities at the statement level. VulChecker \cite{mirskyvulchecker} presents a unique approach, using an intermediate representation called LLVM for vulnerability localization.

\subsection{Code Vulnerability Root Cause}
Identifying the root cause of code vulnerabilities is challenging due to the disconnect between where code crashes and the actual origin of the vulnerability. While the crash occurs at the vulnerability's localization point, the true root cause typically lies in preceding code lines. Over time, several approaches have been developed to address this challenge.
AutoPaG \cite{lin2007autopag} utilized program data flow analysis to detect vulnerabilities such as out-of-bound errors, buffer overflows, and general boundary condition issues. Choi et al. \cite{choi2002isolating} proposed narrowing down differences in thread schedules to isolate the root cause of thread scheduling problems. Failure Sketching \cite{kasikci2015failure} introduced a cooperative approach combining static and dynamic analysis to identify root causes of production failures.
More recent works have focused on automating root cause analysis. ARCUS \cite{yagemann2021arcus} employs execution flags and binary-level analysis, while Yagemann et al. \cite{yagemann2021automated} track execution traces to enhance detection of issues like buffer overflow and use-after-free exploits.
However, a significant gap in current research exists: these methods primarily focus on dynamic analysis and runtime behavior, leaving source code vulnerabilities largely unaddressed. Static vulnerabilities, which can be identified through code inspection without execution, require different approaches for effective root cause analysis. This gap highlights the need for specialized techniques to identify and analyze the root causes of source code vulnerabilities.

\subsection{Explainability Methods for Vulnerability Analysis}
While the aforementioned methods focus on vulnerability analysis during a program's execution, it is equally critical, from a developer's perspective, to scrutinize vulnerabilities during the development phase \cite{islam2023unbiased} \cite{braz2022less}. This has led to the proposal of explainable techniques that pinpoint relevant features contributing to a program's vulnerability \cite{mosolygo2021towards}, \cite{sotgiu2022explainability}, \cite{mao2020explainable}. Asm2Seq \cite{taviss2023asm2seq} and VulANalyzeR \cite{li2023vulanalyzer} took this concept further by introducing explainable deep learning-based approaches for identifying binary vulnerabilities in source code. Notably, VulANalyzeR \cite{li2023vulanalyzer} employed an attention-based explainable mechanism to unearth the root cause of vulnerabilities.


\section{Motivation}
\label{3_Motivation}

As the complexity of software continues to increase \cite{alenezi2020relationship}, the significance of rigorous code security analysis in the deployment stage becomes paramount. Without such measures, we leave our systems at risk, with potential vulnerabilities wide open to threat actors. These actors can exploit backdoors, unintentionally introduced by developers, that remain within the codebase and pose a severe threat to system integrity. Proactively averting such vulnerabilities before deployment by analyzing source code testing constitutes a relatively less laborious approach, resulting in substantial savings in time and reduced economic impact. While there are tools to help developers proactively write secure code, their actual effectiveness in helping developers fix their vulnerable code remains largely unmeasured. Recent open-source static application security analysis tools exhibit limitations in addressing detected vulnerabilities as they fail to provide actionable recommendations and comprehensive explanations supporting their decisions for the developers. This gap underscores the critical need for a comprehensive study to evaluate the effectiveness of these tools in empowering developers, particularly junior ones, to identify and address vulnerabilities proactively.

To inform the development of more effective security tools and practices, we conducted a series of motivation experiments aimed at answering three critical questions: (1) How well can junior developers write secure code without assistance of tools? (2) How well can junior developers write secure code with the assistance of existing software security tools? and (3) What additional features do junior developers desire in security tools to help them fix code vulnerabilities? These experiments not only provide insights into the current state of junior developers' security skills but also highlight the limitations of existing tools and identify potential areas for improvement. In this section, we first detail the demographics, recruitment process, compensation, and ethical approval for our study participants. We then present two experiments examining junior developers' proficiency in writing secure code, both with and without the assistance of tools. Finally, we analyze the results of a questionnaire designed to understand developers' needs and preferences for security tool features, with a particular focus on identifying the root causes of vulnerabilities.

\subsection{Study Participants}

\subsubsection{Participant Demographics}

A total of 23 study participants, all of whom are college students and can be classified as junior developers (defined as those with low experience and limited industry exposure), were involved in this phase of the research. These Computer Science (CS) students at a public research university ranged from juniors in their Bachelor's program to Ph.D. candidates. Each participant completed a demographic questionnaire providing information about their educational background, programming experience, and language preference. Approximately 70\% of the study participants had more than two years of programming experience. However, only one participant had taken security-related courses. The group included 6 junior university students, with the remaining participants (about 74\%) being senior undergraduate or graduate (MS, PhD) students. Almost 60\% of the study participants consisted of graduate students. Given their career paths and goals, these participants represent potential entrants to the software industry within the next one to two years. Consequently, they form an ideal population from which we can gain insights about code security knowledge acquired through the educational system. Further details of the participants are presented in Table \ref{tab:demo_moti}.

\subsubsection{Participant Recruitment}

An email invitation was sent to students in the Computer Science department. The invitation was further expanded via fliers distributed across the campus with the approval and support of the department's admin. In addition, only undergraduate and graduate students who have either completed or were currently enrolled in at least one programming language course were considered. Participants came from diverse educational backgrounds and experiences: 1) Undergraduate Students with programming experience, 2) Graduate Students without industry experience, 3) Graduate Students with industry experience, and 4) Graduate Students with research experience were selected. From our background questionnaire, we found that approximately 41\% of our participants only use print debugging methods, a staggering 82\% of the participants use manual input test cases to test the functionality of their system, and more alarmingly, 94\% of the participants have received no training on code security. We provide a more detailed background on the participants and their demographic information in Appendix Table \ref{tab:demo_moti}, the user study workflow in Appendix \ref{sec:study_workflow}, and the questionnaire with responses to our survey can be found in the code release. Many of these participants seek to join the industry after graduation. Their selection is strategic, considering they possess limited knowledge and training in code security and are ideal candidates for our analysis.

\subsubsection{Participant Compensation and Study Approval}

Each participant received a compensation of US\$25 for their contribution to the study. This study involving human subject participants was approved by our institution’s Institutional Review Board (IRB).

\subsection{Junior Developer Proficiency Without Tools}
\label{sec:without_tools}

To evaluate junior developers' coding proficiency and ability to write secure code without assistance, we designed an experiment centered around a grocery store management system. We crafted 10 function stubs in C, each representing a specific task within the system. Function stubs, also known as method stubs, serve as preliminary models for functions in software development. They define the function's signature but contain minimal or no implementation, allowing for prototyping and outlining program structure.

Each function stub was intentionally designed so participants' responses may fall susceptible to one particular type of vulnerability. The code snippet in Listing \ref{lst:listing-cpp} presents one of ten function stubs participants needed to complete, covering core C programming concepts such as file I/O and memory management. This function takes a long integer parameter; any other form of integer declaration may cause an Integer Overflow Wraparound (CWE-190) error during testing. Similarly, other functions are designed to potentially encounter Null Pointer Dereference (CWE-476), Use After Free (CWE-416), and similar vulnerabilities. The participants were tasked to download the 10 C function stubs. Each function stub included implementation instructions, details of parameter values, details of return values, and usage samples. All participants were required to document each function's start and development completion times. We put details on the CWEs we analyzed in Appendix \ref{cwe_explain}. It is important to note that no code security requirement was requested from the participants at this stage. This was done to observe the participants' unbiased natural coding style and code development quality without the assistance of tools. The study also ensured that the implementations of these functions steered clear of advanced data structures to avoid unnecessary complexity. Essential C libraries and user-defined structures were provided with the function stubs beforehand. Participants were neither encouraged nor discouraged from making use of online resources, including Generative AI or a specific Integrated Development Environment (IDE).

Our assessment consisted of two main components:
\begin{itemize}
    \item \textbf{Functionality Evaluation}: Participants were required to complete each function stub to meet the intended requirements. We employed a set of test cases to verify the functionality of each submitted code sample.
    \item \textbf{Security Analysis}: A second set of test cases was designed to uncover potential vulnerabilities or run-time errors that participants might have inadvertently introduced in their code.
\end{itemize}

Through this experiment, we aimed to quantify the percentage of participants who could write both functional and vulnerability-free code without the assistance of any tools. This approach allowed us to assess not only the basic coding skills of junior developers but also their inherent ability to consider and implement security measures in their code.

\begin{lstlisting}[style=CStyle, label={lst:listing-cpp}, caption=Sample of a function stub aimed to be completed by the participant. The completed function should be able to handle integer overflow and properly set up a base case to avoid a stack overflow if solved recursively.]
long calculateCombinations(long numItems)
{
    /**
     * Problem:
     *  In a grocery store, there are different types of items available for purchase. The store
     * manager wants to calculate the total number of possible combinations of items that a
     * customer can buy. This function should return the factorial of numItems (i.e. numItems!)
     *
    
     * Returns:
     *  the total number of possible combinations.
     */
    //  --> PARTICIPANT: ENTER HERE TIME STARTED <--
    return 0;
}
//  --> PARTICIPANT: ENTER HERE TIME COMPLETED <--
\end{lstlisting}

\begin{table}[t]
\centering
\caption{Success rate of Baseline Group participants on completing ten different function stubs that are functional and are without vulnerabilities.}
\begin{tabular}{l|r|r}

\hline
Function Name                  &   CWE Number &     Success     \\ \hline
\texttt{calculateCombinations} &       119    &      17\%       \\
\texttt{extractPrice}          &       264    &      43\%       \\
\texttt{exportPrices}          &       125    &      22\%       \\
\texttt{*loadPrices}           &       200    &      25\%       \\
\texttt{printMaxPrice}         &       416    &      47\%       \\
\texttt{validateUserCreation}  &       399    &      30\%       \\
\texttt{addUser}               &       20     &      19\%       \\
\texttt{removeUser}            &       476    &      26\%       \\
\texttt{promptUserCreation}    &       189    &      33\%       \\
\texttt{*resizeDatabase}       &       190    &      27\%       \\ \hline
\end{tabular}
\label{tab:moti_1}
\end{table}

The results of this experiment are presented in Table \ref{tab:moti_1}. We define the success rate as the percentage of participants who wrote a functional code without introducing any vulnerabilities. Out of the ten functions completed by each of the 23 participants, we observed varying levels of success across different Common Weakness Enumeration (CWE) categories. The functions associated with CWE-119 and CWE-20 had the lowest success rates at 20\% and 18\%, respectively. These functions involved writing data outside the defined memory buffer and performing improper input validation. The remaining functions showed slightly higher success rates, with CWE-416 reaching 46\%. This low performance in writing functional and secure code, particularly for memory management and input validation, suggests that participants may be insufficiently aware of or attentive to crucial security considerations when coding these types of functions. The results show that these junior developers struggle with writing secure and functional code.

\subsection{Junior Developer Proficiency in Fixing Vulnerable Code With Existing Tools}
\label{sec:motivation_existing_tool_effectiveness}

The second phase of our study aimed to assess junior developers' ability to fix vulnerable code using information provided by state-of-the-art (SOTA) source code vulnerability analysis tools. We considered various popular tools and techniques, including Infer \cite{infer}, Cppcheck \cite{cppcheck}, and deep learning methods for vulnerability detection \cite{zhou2019devign} \cite{islam2023unbiased}, localization \cite{fu2022linevul} \cite{pornprasit2022deeplinedp} \cite{mirskyvulchecker}, and repair \cite{fu2022vulrepair} \cite{pearce2023examining}.
These tools provide valuable information for classifying and localizing vulnerabilities, and are not designed as interactive assistive tools to help developers understand detected vulnerabilities. The goal of such tools should be to educate developers with appropriate security knowledge, helping them become more self-sufficient in writing secure code, rather than completely offloading the task.
To evaluate the efficacy of SOTA techniques in assisting developers, we conducted a second coding exercise. Participants were presented with five new vulnerable code functions and provided with information similar to that given by SOTA techniques such as VELVET \cite{ding2021velvet}, VulChecker \cite{mirskyvulchecker}, and LineVul \cite{fu2022linevul}. Specifically, they received: 1) The Common Weakness Enumeration (CWE) classification of the detected vulnerability, and 2) The line(s) of code where the vulnerability was found.

\begin{figure}[t]
        \centering
        \includegraphics[width=0.48\textwidth]{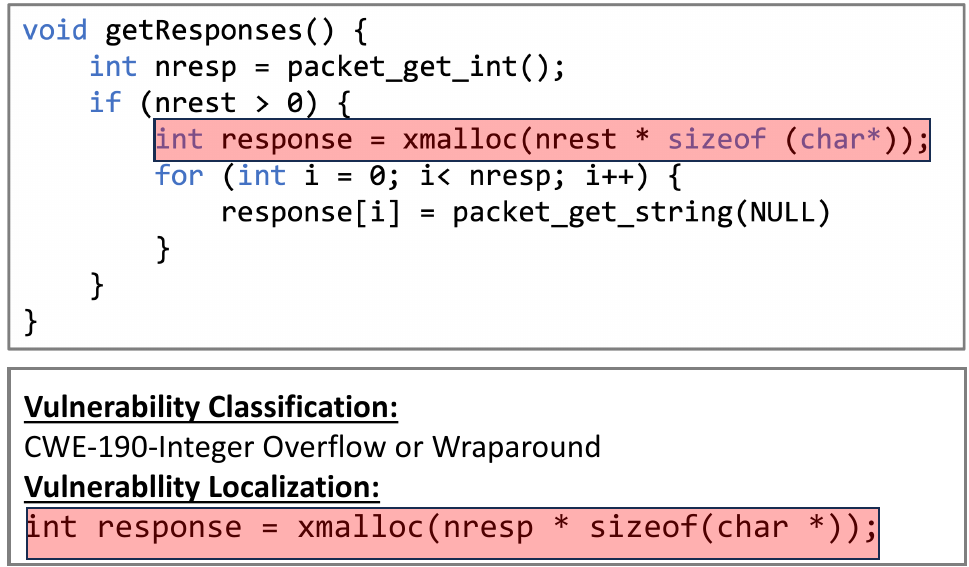}
    \caption{Sample source code provided to the participants depicted at the top and output at the bottom provided by the SOTA techniques. We conducted our initial survey by providing the participants with this information and determined their capability to repair vulnerability using these two outputs: classification and vulnerable line.}
    \label{fig:survey_sota}
\end{figure}

To present this information to the participants, we first processed each of the five vulnerable functions using VELVET, VulChecker, and LineVul to obtain CWE classifications and vulnerable line identifications. In cases where the outputs of these three models conflicted, we used the majority outcome to ensure consistency and minimize false positives or negatives. Participants were then tasked with repairing the vulnerability in each function based on the provided information.
After all submissions were collected, we calculated the percentage of participants who successfully repaired each vulnerable code. A sample of the information given to participants to aid in their repair of vulnerable code is presented in Figure \ref{fig:survey_sota}.

The results presented in Table \ref{tab:moti_2} reveal that the majority of participants struggled to fix the given vulnerable code functions, even with the assistance of information provided by SOTA techniques. While the classification and localization data offered some insight into the code vulnerabilities, the success rate in fixing these issues remained below 50\%. This performance, although an improvement compared to writing code without any suggestions or assistance, still leaves a significant portion of the code potentially vulnerable. In a real-world scenario, this outcome suggests that approximately half of the code could remain susceptible to security threats, even when developers are aided by current SOTA tools. These findings align with recent research that emphasizes the need for more comprehensive security support and training for software developers \cite{green2016developers, acar2016you}. 

\subsection{Participant Questionnaire}

To contextualize our findings and explore potential improvements to security tools, we conducted a structured questionnaire following participants' code writing exercises without and with source code vulnerability analysis tool assistance. This survey served two primary purposes: first, to evaluate participants' understanding of security concepts and their approach to problem-solving, and second, to gather insights on what features could be added to security tools to better assist developers in fixing vulnerable code. The questions were designed to provide a comprehensive view of the participants' experience and needs in the context of code security.

Our findings revealed that only 42.7\% of participants were familiar with Common Weakness Enumeration (CWE) vulnerability categories prior to the coding exercises. When asked about their approach during the exercises, 66.6\% of participants reported occasionally consulting the Internet for assistance, while 27.8\% admitted to extensively using it to complete the tasks. Notably, 37\% of participants disclosed that they utilized ChatGPT to help them with their responses during the coding challenges. This indicates not only a lack of prior knowledge about security vulnerabilities but also a significant reliance on external resources, including AI tools.

When asked to rate their comprehension of the vulnerabilities based on the provided code snippets, classification, and localization information on a scale of 1 to 5, the results were revealing. Only 7.1\% of participants reported a clear understanding (5/5), while the majority (42.9\%) rated their understanding at a moderate level (3/5). These results highlight that despite access to various resources, many participants still struggled to fully grasp the nature of the vulnerabilities they encountered.

In addition to assessing participants' knowledge and resource usage, we sought to gather insights on what features could be added to security tools to better assist developers in fixing vulnerable code. Some indicated they would have benefited from clearer definitions of the CWE vulnerabilities they encountered, suggesting that current tools may assume a level of background knowledge that not all junior developers possess and that these classifications don't necessarily help them secure their code. Notably, about 22\% of participants requested information relating to finding the root cause of the vulnerabilities. This feedback suggests that while current tools can identify and localize vulnerabilities, they often fall short in providing the contextual understanding necessary for developers to effectively address the underlying issues.

\begin{table}[t]
\centering
\caption{The success rate of participants from the control group who could repair the function only when the vulnerable line (vulnerability localization) and the CWE class (vulnerability classification) of the vulnerability for each function were provided using existing static code vulnerability analysis tools.}
\begin{tabular}{l|r|r}

\hline
Function Name               &      CWE-Number     & Success \\ \hline
\texttt{getValueFromList}   &      125        &      47\%      \\
\texttt{*callHelper}        &      416        &      48\%      \\
\texttt{SQLConnect}         &      264        &      20\%      \\
\texttt{readFile}           &      416        &      20\%      \\
\texttt{*createBoard}       &      20         &      46\%      \\ \hline
\end{tabular}
\label{tab:moti_2}
\end{table}


\section{Threat Model}
\label{4_preliminaries}
We start the definition of our threat model with the assumption that rule or AI-based tools can be used to offload the task of vulnerability repair. Given this context, two types of threat actors must be considered in the software development landscape. In the first category we have human developers writing vulnerable code due to a lack of knowledge in code security. The second threat actor category involves assistive platforms like ChatGPT, GitHub, or StackOverflow inquired by many developer to reuse code that could be potentially insecure. 

Our threat model primarily considers developers working with vulnerable C/C++ code. Our preliminary user study revealed that software developers inadvertently introduce diverse vulnerabilities into their code. These vulnerabilities may compromise the physical memory and CPU caches thereby allowing threat actors to gain privileged root mode access. Attackers could exploit such vulnerabilities through various attack scenarios, like buffer overflow, code injection, improper operations within a memory buffer, or similar vulnerabilities related to these. These vulnerabilities can subsequently provide control over the system, enable data theft, or even launch further attacks. Our approach to minimizing vulnerability in source code is two-fold. Firstly, we propose a system to assist developers in 1) detecting a code vulnerability, 2) classifying it (CWE category), 3) localizing vulnerable lines, and 4) finding the root cause of the code vulnerability. Secondly, with the combination of these pieces of information, we aim to provide useful security information to assist developers in writing secure code.

\section{Proposed System Architecture}
\label{5_proposed_system}
\begin{figure*}[h]
    \centering
    \includegraphics[scale=0.54]{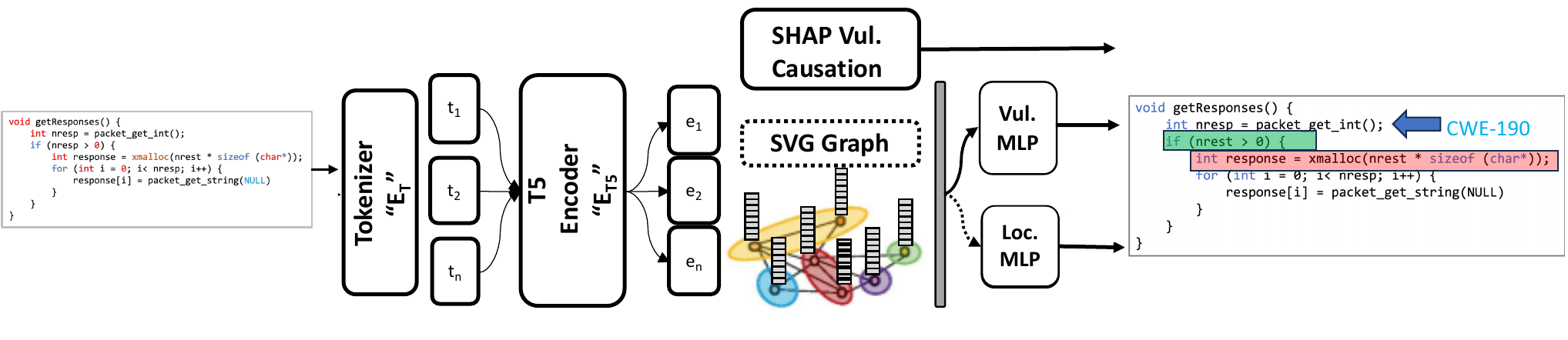}
    \caption{The proposed architecture of our system: An LLM-powered source code diagnostic tool that assists programmers in vulnerability identification, classification, localization, and the root cause of vulnerability}
    \label{fig:fig_2}
\end{figure*}

We provide an end-to-end system to analyze source code vulnerabilities and demonstrate our system's capability to assist developers in writing secure and vulnerability-free code. Figure \ref{fig:fig_2} depicts the overall architecture of our proposed vulnerability resolution and evaluation procedure. Our proposed contribution is divided into three steps: i) Vulnerability Classification and ii) Vulnerability Localization using Supervised Fine Tuning with a T5-RCGCN model, and finally, iii) Vulnerability root cause analysis using explainable techniques.

\subsection{Problem Formulation}

To assist developers in analyzing and fixing vulnerable code and simultaneously improving their skills to prevent introducing future vulnerabilities, it is essential to address the problem of finding the root cause of the code vulnerability. This source code needs to be converted into a different structure so that it can be ingested by an AI model. Each program function, denoted as $p_i$, is converted into a multi-edged graph, $\mathcal{G}_{raph}$. This graph is constructed such that the set of nodes, $T$, represent the programming language tokens, and the edges, $E_{dge}$, indicate the connections between these nodes. We transform the edge pairs into an adjacency matrix $A$.

From this structure, the model should be capable of detecting and classifying a vulnerability into the correct category, denoted as $CWE$, that a vulnerable input function $p_i$ aligns with. After classifying $p_i$, the model should also localize the vulnerable lines. The process is done by finding starting ($L_{start}$) and the ending ($L_{end}$) line pair, $[L_{start}, L_{end}]$ which is vulnerable. In the context of our system, we define three outputs: the vulnerability identification with classification, $CWE$, the vulnerability localization range $[L_{start}, L_{end}]$, and the root cause of vulnerability $V_{root}$.

\textbf{Root Cause Vulnerability Identification through Explainability:}
The proposed architecture uses explainable techniques to discern the impacts contributed by each program token, denoted $t_i \in T$, on the vulnerability prediction of a given program.

This model, initially trained to classify and localize the vulnerabilities, garners a holistic understanding of the vulnerabilities and correspondingly attributes weights to each token in the source code based on their association with vulnerability. Moreover, our model proficiently comprehends contributions from individual tokens based on their attribution weights. By prioritizing specific tokens over others, the model inherently puts more significance on the critical tokens that show a higher likelihood of vulnerability.

For the scope of this research, the weight associated with a token $t_i$ is symbolized as $\phi\textsubscript{i}$. The predictive output delivered by our model for one program is mathematically defined as follows:

\begin{equation}
\hat{A} = f({T}) = \phi \textsubscript{0} + \sum_{i=1}^{N} \phi \textsubscript{i} {t_i}
\end{equation}

Where $\hat{A}$ is the set of attributions for each token in $T$, and $N$ is the total number of tokens. In this expression, the weight $\phi\textsubscript{i}$ encapsulates the contribution of a token $t_i$ to the model's overall output.

\textbf{Assessing System Efficacy:}
To address the issues we found in our initial study in our motivation study in Section \ref{3_Motivation}, we aim to answer the following three Research Questions (RQs):

\noindent\textbf{RQ1:} \textit{How efficiently can we classify and localize the vulnerability in addition to finding the root cause vulnerability of source codes?}

\noindent To find how efficiently we can localize vulnerability, we use a metric called IoU. Furthermore, in order to measure the classification accuracy, we measure F1 Accuracy scores. The higher classification and localization performance will ensure the root cause analysis's effectiveness.

\noindent\textbf{RQ2:} \textit{Does our root cause detection system generalize enough to identify zero- and $n$-day vulnerabilities from the wild?}

\noindent To answer this question, we analyzed several open-source projects written in C/C++, and our internal security experts manually checked the validity of the root cause provided by our system.

\noindent\textbf{RQ3:} \textit{Using the root cause of vulnerability, how effectively is our system assisting software developers in fixing code and educating developers in writing secure code with fewer vulnerabilities?}

\noindent To find the root cause of source code vulnerability, we used an explainability-based technique to determine the importance of tokens. Then we propose an in-depth survey analysis to determine how effective our system is compared to the current SOTA methods.

These RQs serve as a guideline to evaluate our system's vulnerability detection and localization capabilities, its competency in providing actionable insights for vulnerability remediation, and its adaptability in discerning new vulnerabilities across diverse environments.

\subsection{Code Vulnerability Detection and\\ Classification}

For code vulnerability detection and classification, the input source code has to go through pre-processing steps and, finally, through our proposed T5-RCGCN.

\paragraph{\textbf{Source Code Representation}} In this pre-processing step, the input is an entire function of source code, which may be vulnerable or non-vulnerable. We initially employ the CodeT5 \cite{wang2021codet5} tokenizer, which tokenized words using a byte-pair fashion \cite{sennrich-etal-2016-neural}. CodeT5 tokenizer was pretrained in programming languages like C/C++ to extract the set of tokens $T$ from a given function $p_i$.

We analyzed the individual functions by random sampling from our datasets and found that the average number of tokens is approximately 490. Therefore, we propose to select 512 as the maximum number of tokens, and we trim the length of the set of tokens $T$ to 512. Moreover, we add two unique tokens, $<BOS>$ and $<EOS>$, at the beginning and end of the program as a separator. If the length of the program is less than 512, we use a unique token $<PAD>$ to resize the length to 512. After finalizing the nodes, we develop $\mathcal{G}$ by connecting the nodes using SVG \cite{islam2023unbiased}.

\paragraph{\textbf{Source Code Semantic Graph Representation}} We have refined the process of root cause analysis within source code by enhancing the capabilities of Graph Convolutional Networks (GCN) through the incorporation of a Semantic Vulnerability Graph (SVG) \cite{islam2023unbiased}. The SVG combines four distinct categories of edges, encompassing data \cite{zhou2019devign}, control \cite{zhou2019devign}, sequential \cite{huang2019text, zhang2020every}, and poacher flow \cite{islam2023unbiased} relationships. These four edge categories comprehensively capture the source code's syntactic and semantic attributes.

By combining these diverse graph types, the GCN gains an intricate understanding of the source code, enabling a contextual interpretation and passive runtime understanding \cite{islam2023unbiased} of the source code. This contextualization facilitates the creation of relational representations for different program tokens, significantly enhancing the system's ability to pinpoint the tokens responsible for underlying vulnerabilities. This holistic approach to vulnerability analysis represents a significant advancement in source code security analysis.

\paragraph{\textbf{CodeT5 Encoder}}
Our proposed system is powered by CodeT5 \cite{wang2021codet5}, a large language model that adopts the encoder-decoder architecture inspired by T5 \cite{2020t5}. It effectively captures the syntactic structure of code and utilizes positional information associated with each token to facilitate token-based localization.

We use the encoder of CodeT5, which consists of multiple layers of self-attention and feed-forward neural networks, to generate the embedding of each node or token in our graph. The self-attention mechanism computes attention weights to capture the input sequence's interdependencies and relationships between elements. This allows for encoding contextual information from the nearby tokens in the code. The output from the self-attention layer is then passed through a feed-forward neural network, which applies a nonlinear transformation independently at each position and finally outputs an embedding vector $E$ of size 768 for each token. This embedding vector acts as the node representation for each token, which is then converted into an adjacency matrix using SVG \cite{islam2023unbiased} and passed to the GCN layer.

\paragraph{\textbf{T5-RCGCN}} Graph Convolution Network (GCN) attempts to comprehend the correlation between any pair of node embeddings of tokens from code we got from the CodeT5 encoder. We introduce a two-layered GCN with a residual connection. Mathematically, we implemented GCN as follows:

\begin{equation}
\label{eqn:7}
F_{GCN} = H\textsuperscript{(n + 1)} = H\textsuperscript{n} + \sigma \biggl( W^n_{GCN} H\textsuperscript{n} A \biggr )
\end{equation}

Here, $W^n_{GCN}$ represents the learnable weights at the $n$-th layer, and $H\textsuperscript{n}$ is the feature representation of all tokens $T$ from a function $f_i$ at the $n$-th layer. $H\textsuperscript{(0)} = E$ and $A$ represents the adjacency matrix. The multiplication of the matrices $W^n_{GCN}$, $H_n$, and $A$ is followed by an activation function $sigma$ (e.g., $ReLU$). $F_{GCN}$ is the final representation generated by our proposed GCN.

\paragraph{\textbf{Loss Function}} We use the Focal Loss function \cite{lin2017focal}, built on top of cross-entropy, which can handle possible data imbalance issues as identified by \cite{islam2023unbiased} for vulnerability classification purposes. Our Focal Loss function stands thus:

\begin{equation}
    \label{eqn:loss1}   
FocalLoss(p_{rob}^t) = -\alpha (1 - p_{rob}^t)^\delta \log(p_{rob}^t)
\end{equation}

In this instance, $\alpha$ denotes the balancing factor between the number of vulnerable and non-vulnerable code samples, while $p_{rob}^t$ is the probability distribution of our model's output. We use $\delta$ as an adjustable parameter that distinguishes between easy and hard examples \cite{lin2017focal}.

\paragraph{\textbf{Detection and Classification}}For vulnerability detection and classification purposes, we use the feature vector $F_{GCN}$, produced by our proposed T5-RCGCN. We added a dense layer after the feature vector layer generated by GCN. The dense layer \texttt{Vul. MLP} generates the CWE number of the vulnerable code if a vulnerability exists, as depicted in Figure \ref{fig:fig_2}. If no vulnerability exists, the \textit{Vul. MLP} layer generates an output of 0. Furthermore, for each identified vulnerability classified by a CWE number, we provide a static description of the identified vulnerability.

\subsection{Identification of Vulnerable Lines}
In order to find the vulnerable lines, our proposed model identifies a block of code by generating the starting and ending lines of the vulnerable code. The second dense layer \texttt{Loc. MLP} generates $L_{start}$ and $L_{end}$, the line range where the vulnerability exists. Therefore, we connect $F_{GCN}$ with another dense layer \textit{Loc. MLP} for finding the vulnerable line. 

Since line numbers vary depending on the position of the vulnerable line in code, we designed the identification of vulnerable lines as a regression problem. Hence, we apply Mean Squared Error (MSE) loss for vulnerability localization. Our MSE loss function is defined as follows:

\begin{equation}
    \label{eqn:loss2} 
MSE = \frac{1}{n} \sum_{}^{} (L - \hat{L})^2
\end{equation}

where $L$ is the original outcome and $\hat{L}$ is the outcome from the model.

\subsection{Root Cause of Vulnerability}

After the model is sufficiently trained to classify and localize the vulnerability, we find the root cause of the vulnerability using our trained model. We employed DeepLiftSHAP attribution scores. We hypothesize that, since the model can effectively classify and localize the vulnerability,  we determine to utilize the model's understanding of vulnerability to determine the contribution of each token to find the root cause of vulnerability.

\begin{algorithm}[t]
    \caption{Token Attribution for Root Cause Vulnerability}
    \label{alg:algorithm}
    \textbf{Input}: $model$,  input program $p_i$ \\
    \textbf{Output}: Explainable attribution scores $\hat{A}$ for tokens $T$  

    \begin{algorithmic}[1] 
        \STATE $T$ = $Tokenizer (p_i)$
        
        \STATE $\hat{A}$ = [$t_i$: 0 for $t_i$ in $T$] \\
        
        \STATE $original_{pred}$ = $model(T)$

        \STATE $contrib_{subset}$ = $[]$
        \FOR{each $t_i$ in T}
            \STATE $contribution$ = DeepLIFT($t_i$)
            \STATE $contrib_{subset}.append(contributions)$

        \ENDFOR
        

        \FOR{Each $s_{sub}$ in $contrib_{subset}$}
            
            \STATE $subset_{pred}$ = $model(s_{sub})$

            \STATE $marginal_{contr}$ = $original_{pred}$ - $subset_{pred}$

            \STATE $norm_{contr}$ = $marginal_{contr}$ / $len(s_{sub})$

            \STATE $\hat{A}.append(norm_{contr})$
        \ENDFOR

        \STATE \textbf{return} $\hat{A}$
    
    \end{algorithmic}
\end{algorithm}

DeefLiftSHAP is an explainability technique for neural networks based on executing a SHAPly \cite{lundberg2017unified} variant of the original DeepLift \cite{shrikumar2017learning}. Combining DeepLIFT and SHAPly, DeepLiftSHAP operates on deep learning frameworks to explain neural network models. We generate attribution scores based on the DeepLiftShap \cite{lundberg2017unified}, where we generate the code token attribution scores based on our proposed Algorithm \ref{alg:algorithm}. We sum up the attribution scores of each token in a line to generate an attribution score for each line. Here, $\hat{A}$ is the set of attribution scores for all tokens in $T$ of a function $p_i$, where, $\hat{A} \in \{a_1, a_2, ... , a_m \}$. After generating scores for each line or statement, we consider the line with the highest attribution values before $V_{Start}$ as the root cause of the vulnerability.

\section{System Implementation and Evaluation}
\label{6_experiments_discussion}

\subsection{Experimental Datasets}
\label{sec:5_dataset}

\textbf{D2A and BigVul} As per our objective to provide an end-to-end solution for vulnerability management, we made use of datasets containing source code from real-world applications. Big-Vul \cite{fan2020ac} and D2A \cite{zheng2021d2a} were used for vulnerability classification, localization, and root cause analysis. BigVul provides ten vulnerability categories that fall within the top 25 CWE vulnerabilities mentioned at CWE \cite{cwe}. D2A contains open-source projects from GitHub, and labels were created using commit filtering and static analyzer tools. The BigVul dataset provides CWE numbers of vulnerabilities, while D2A does not provide a vulnerability class, so we use it only for vulnerability detection and localization. D2A consists of 6,728 samples classified into two classes: Vulnerable and Non-Vulnerable. BigVul, on the other hand, contains 217,007 samples categorized into ten CWE classes: 119, 20, 125, 200, 264, 399, 416, 476, 189, and 190.

\textbf{IoT OS Repositories}
We collected a dataset from six OS repositories from GitHub to test the capability of our system in finding N-day and zero-day program samples in real-world operational systems. Six IoT operating repositories were downloaded to measure this metric: TinyOS, Contiki, Zephyr, FreeRTOS, RIOT-OS, and Raspberry Pi OS. The entire repository of these operating systems was scanned using JOERN \cite{joern}, a tool specially designed to monitor and analyze large repositories. JOERN command line interface was used to split the C/C++ files into functions for operational analysis. This process yielded a substantial number of function definitions across the repositories: TinyOS provided 13,650 functions, Contiki 14,272, FreeRTOS 88,033, RIOT-OS 37,458, Zephyr 44,621, and Raspberry Pi OS, the largest in the set, contributed 267,623 functions.

\subsection{Evaluation Metrics}

Standard metrics in this space were used for our quantitative analysis. For vulnerability classification purposes, the standard metrics including accuracy, precision, recall, and F1 score metrics were used. Performance in vulnerability localization is measured by establishing a boundary between vulnerable lines (starting and ending vulnerable lines). Given these limits, the metric Intersection of Union (IoU) is employed. Since our input data consists of source code, which is linear single-dimensional data, unlike a 2D image, the IoU formula \cite{rezatofighi2019generalized} was modified to 1D scale. Let's consider our model predicts the localization line boundaries from $ \hat{Vul_{Code}} = [\hat{L_{Start}} - \hat{L_{End}}]$ and $Start <= End$ and the ground truth for localization is $Vul_{Code} = (L_{Start}, L_{End})$. For our purposes, the IoU is:

\begin{equation}
\label{eqn:11}    
IoU = \frac{|\hat{Vul_{Code}} \cap Vul_{Code}|}{|\hat{Vul_{Code}} \cup Vul_{Code}|}
\end{equation}

If the value of IoU is zero, Equation \ref{eqn:11} shows that there is no overlap between $\hat{L}$ and $L$, indicating that the model was not able to locate a single vulnerable line. On the other hand if the value is 1 it stated that the model detected all vulnerable lines. 

\subsection{Experimental Setup and Implementation}

The datasets were split into 80:10:10 ratios for training, validation, and testing in these experiments. A 12-layer CodeT5 encoder was used to generate the embeddings, and a two-layer GCN with a residual connection was used to generate feature vectors for each function $p_i$. The final feature representation vector produced by GCN is of size 512. The generated model was trained for 20 epochs, a maximum token length of 512 was set for each function processed by the model, and a learning rate of 6e-6 was set. Eight A100 NVIDIA GPUs were used to train our proposed model.

As presented in Figure \ref{fig:fig_2}, the model learns the classification and localization tasks during the training process. Cross-entropy loss was used for vulnerability classification, and MSE loss was used for localization purposes. Along with the CWE Number provided by the vulnerability classification feature, a description associated with the CWE Number is provided to the developers.

We added an output layer with ten neurons to classify ten vulnerability classes. In addition, vulnerability localization is provided by stating a line range of the statements where the vulnerability exists. This is provided by a separate output layer composed of two units, where one states the starting line number for the first and the second states the last line of the vulnerable statement. Finally, to provide an analysis of the root cause of the vulnerability, an explainability technique using DeepLiftSHAP attribution scores was employed on the trained model.


\subsection{Evaluation Overview}
We first present the results of T5-RCGCN on vulnerability classification and vulnerability localization by training and testing on two common source code vulnerability datasets. We then test the generalizability of our approach, by training on one dataset and testing on another dataset. Next, we do n-day and zero-day testing to see if our system can detect these vulnerabilities in open source IoT repositories. These tests demonstrate our system's robustness and establish T5-RCGCN's state-of-the-art performance in vulnerability classification and localization across diverse scenarios. We then show how using DeepLiftSHAP to additionally identify the root cause of the code vulnerability impacts junior developer’s ability to secure vulnerable code. Finally, we compare how much junior developers learn how to write secure code after having used T5-RCGCN vs other source code analysis tools to see if they have improved their understanding on writing secure code without assistance.

\subsection{Vulnerability Classification and Localization Evaluations}

\subsubsection{In-dataset testing}

\begin{table}[t]
\centering
\caption{Vulnerability classification and localization on the Big-Vul and D2A dataset}
\begin{tabular}{p{0.025\textwidth}
                 p{0.08\textwidth}
                 p{0.030\textwidth}
                 p{0.006\textwidth}
                 p{0.006\textwidth}
                 p{0.006\textwidth}
                 p{0.006\textwidth}}

\toprule
\textbf{Data}                    & \textbf{Model}          & \multicolumn{1}{l}{\textbf{IoU}} &  \multicolumn{1}{l}{\textbf{Acc.}} & \multicolumn{1}{l}{\textbf{F1}} & \multicolumn{1}{l}{\textbf{Pre.}} & \multicolumn{1}{l}{\textbf{Rec.}} \\ \midrule
\multirow{4}{*}{\rotatebox{90}{D2A}}    & Devign         & 0.58                                                 & -                        & -                      & -                        & -                        \\
                        & VELVET         & 0.55                                               & 0.59                     & 0.58                   & \textbf{0.70}            & 0.50                     \\
                        & LineVul        & 0.42                                               & -                        & -                      & -                        & -                        \\
                        & PFGCN & -                & 0.61                         & 0.61            & 0.61          & 0.62                           \\ 
                        
                        & \textbf{Ours} & \textbf{0.72}                                      & \textbf{0.62}            & \textbf{0.68}          & 0.60                     & \textbf{0.65}            \\
                        \midrule

\multirow{5}{*}{\rotatebox{90}{BigVul}} & VulChecker         & 0.44                                                 & -                        & 0.26                   & 0.18                     & 0.52                     \\
                        & VELVET         & 0.45                                                 & -                        & -                      & -                        & -                        \\
                        & LineVul        & 0.45                                                 & -                        & 0.56                   & \textbf{0.66}            & 0.60                     \\
                        & IVDetect       & -                                                       & -                        & 0.35                   & 0.23                     & 0.72                     \\
                        & \textbf{Ours} & \textbf{0.49}                                         & \textbf{0.65}            & \textbf{0.62}          & 0.62                     & \textbf{0.66}            \\ \bottomrule
\end{tabular}
\label{tab:class}
\end{table}

We first evaluate our proposed source code vulnerability analysis tool on the standard tasks of vulnerability classification and vulnerability localization by training T5-RCGCN models on the D2A and BigVul datasets. For the D2A dataset, we compared our model with Devign \cite{ding2021velvet}, VELVET \cite{ding2021velvet}, and LineVul \cite{fu2022linevul} and PFGCN \cite{islam2023unbiased}. For the Big-Vul dataset, we compared our model with VulChecker \cite{mirskyvulchecker}, VELVET \cite{ding2021velvet}, LineVul \cite{fu2022linevul}, and IVDetect \cite{li2021vulnerability}.

We present the results of this experiment in Table \ref{tab:class}. On the task of vulnerability classification on the D2A dataset, T5-RCGCN outperforms the next best performing baseline, PFGCN, in F1 score by 0.07 points. Then on the BigVul dataset, T5-RCGCN outperforms the next best performing baseline, in F1 score by 0.06 points.

\begin{figure}[b]
        \centering
        
        \includegraphics[width=0.5\textwidth]{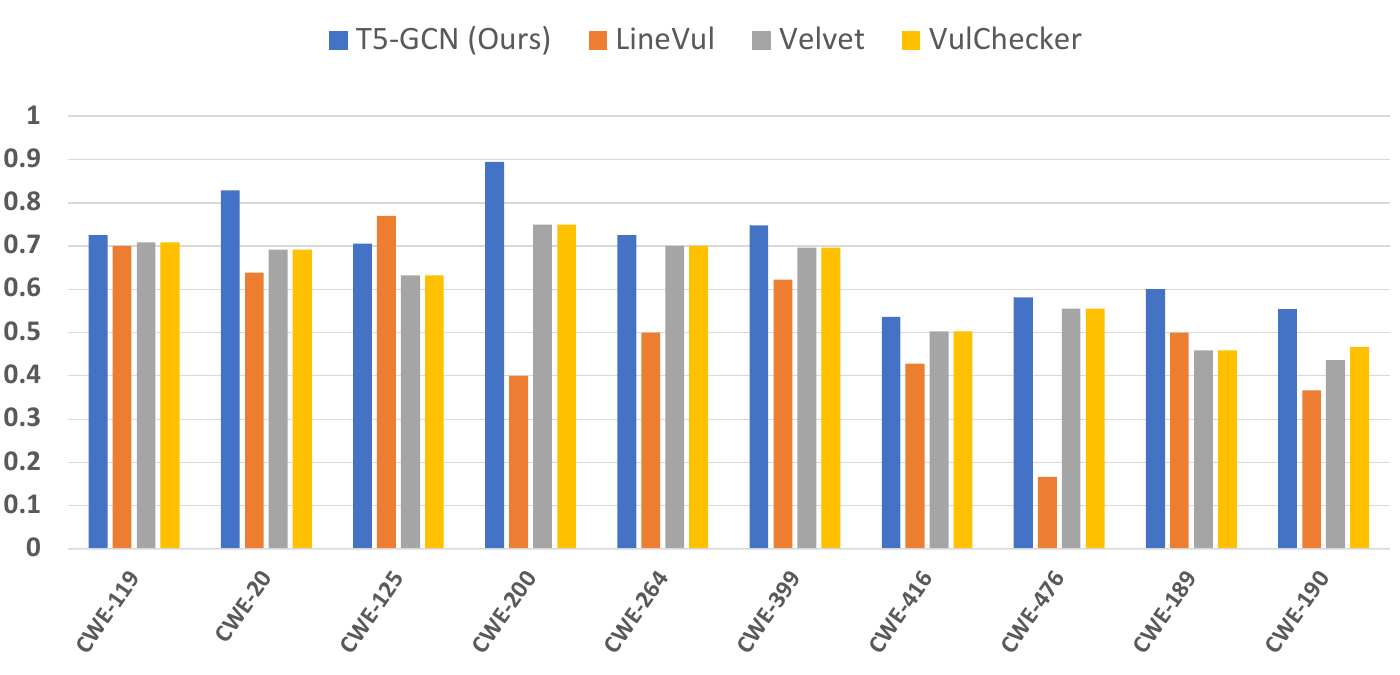}
        
    
    \caption{Multi-class vulnerability classification of vulnerable code from the BigVul datasets, comparing SOTA techniques to our proposed T5-RCGCN. X-axis is the vulnerability category, and Y-axis demonstrates the F1 Score}
    \label{fig:mult_chart}
\end{figure}

We further break down the results of vulnerability classification on the BigVul dataset in Figure \ref{fig:mult_chart}, comparing our method to three other top-performing approaches for the 10 most common categories of vulnerabilities in CWE \cite{cwe} \cite{fan2020ac}.
Our analysis of the training dataset revealed that CWE-119, CWE-20, and CWE-200 have the highest number of training samples, each with over 1000 examples. In contrast, CWE-416, CWE-476, and CWE-190 have the least, with fewer than 400 training samples each. This distribution is reflected in our model's performance, with CWE-119 and CWE-20 achieving F1 scores of 70\% and 63.83\%, respectively. Our model's classification capability significantly outperforms other state-of-the-art models such as VulChecker \cite{mirskyvulchecker}, VELVET \cite{ding2021velvet}, LineVul \cite{fu2022linevul}, and PFGCN \cite{islam2023unbiased}.
For categories with fewer training examples (CWE-416, CWE-476, and CWE-190), we observe a moderate decrease in performance. However, our model still achieves higher performance than SOTA models in these categories, which we attribute to the use of Focal Loss during the training phase. Focal Loss likely helped by addressing the class imbalance problem, giving more weight to the underrepresented vulnerability classes. 

On the task of vulnerability localization, we see that our proposed method also outperforms the other top performing methods on the IoU metric. T5-RCGCN outperforms Devign by 0.14 points on the D2A dataset, and LineVul and Velvet by 0.04 points on the BigVul dataset. Overall, these experiments show that T5-RCGCN outperforms other top methods on vulnerability classification and vulnerability localization on in-distribution data.

\subsubsection{Cross-dataset testing}

\begin{table}[b]
\centering
\caption{Generalizability Testing of our Model. We test with out-of-sample data during evaluation to test the generalizability of the model}
\label{tab:generaility}
\begin{tabular}{p{0.08\textwidth}
                 p{0.07\textwidth}
                 p{0.03\textwidth}
                 p{0.03\textwidth}
                 p{0.03\textwidth}
                 p{0.03\textwidth}
                 p{0.04\textwidth}}
\toprule
\textbf{Training Data} & \textbf{Evaluation Data} & \textbf{IoU} &\textbf{Acc.} & \textbf{F1} & \textbf{Pre.} & \textbf{Rec.}      \\ \midrule

\multirow{2}{*}{D2A} & D2A & 0.72 & 0.62  & 0.68  &0.68  & 0.65   \\
                                    & BigVul & 0.40 & 0.52  & 0.55  &0.57  & 0.52 \\ \midrule
                                                    
\multirow{2}{*}{BigVul} & BigVul & 0.49 & 0.65  & 0.62  &  0.66  & 0.66     \\
                                    & D2A & 0.44 & 0.57  & 0.59  &0.60  & 0.51 \\ \midrule
                                    
\multirow{3}{*}{Combined} & D2A & 0.74 & 0.66  & 0.75  &0.70  & 0.70     \\

& BigVul & 0.50 & 0.64  & 0.66  &0.67  & 0.70 \\

& Combined & 0.65 & 0.68  & 0.64  &0.66  & 0.69 
                                    
\\\bottomrule

\end{tabular}
\end{table}

We next evaluate our T5-RCGCN model's generalization capabilities through cross-dataset testing, where we train on one dataset and test on another. We conducted three experiments: (1) training on D2A and testing on BigVul, (2) training on BigVul and testing on D2A, and (3) training on a combined dataset of D2A and BigVul and testing on each dataset separately.
Table \ref{tab:generaility} presents the results of these experiments. When training on D2A and testing on BigVul, and vice versa, we observe a decrease in performance compared to the in-distribution results reported in Table \ref{tab:class}. This decline in accuracy suggests that the model faces challenges in generalizing across different datasets.
However, when we trained the model on the combined D2A and BigVul dataset, we observed significant improvements in performance. For the D2A dataset, the F1 score increased by up to 7 percentage points, and the IoU score improved by 2 percentage points. The BigVul dataset also saw a 1 percentage point increase in the IoU score.
The improvement in performance for the D2A dataset when using the combined training set is particularly noteworthy. We attribute this to the fact that the D2A dataset is approximately 26 times smaller than BigVul. By combining the datasets, we effectively increase the amount of training data available for D2A, leading to enhanced generalizability of our model.

\subsubsection{N-day and Zero-day testing}

\begin{table}[t]
\centering
\caption{N-day Vulnerability Discovery from IoT OS Repositories. We show that T5-RCGCN system discovered more n-day vulnerabilities than other tools.}
\begin{tabular}{l|r|r|r|r}
\hline
\textbf{IoT OS}                  & \textbf{N-Day Vulnerability} & \textbf{T5-RCGCN} & \textbf{Devign} & \textbf{ReVeAL} \\ \hline
TinyOS                           & N/A                          & 0               & 0               & 0               \\ \hline
\multirow{2}{*}{Contiki}         & CWE-119                      & 4               & 2               & 1               \\
                                 & CWE-189                      & 1               & 1               & 1               \\ \hline
\multirow{3}{*}{Zephyr}          & CWE-264                      & 2               & 2               & 1               \\
                                 & CWE-119                      & 4               & 1               & 2               \\
                                 & CWE-399                      & 1               & 1               & 1               \\ \hline
\multirow{2}{*}{FreeRTOS}        & CWE-119                      & 1               & 1               & 0               \\
                                 & CWE-190                      & 2               & 0               & 0               \\ \hline
\multirow{2}{*}{RIOT-OS}         & CWE-119                      & 5               & 3               & 1               \\
                                 & CWE-476                      & 1               & 1               & 0               \\ \hline
\multirow{2}{*}{Raspberry-Pi OS} & CWE-200                      & 2               & 1               & 1               \\
                                 & CWE-119                      & 1               & 1               & 0               \\ \hline
\textbf{Total}                   &                            & 24              & 14              & 8              \\ \hline
\end{tabular}
\label{tab:iot_data}
\end{table}

We evaluate our model's ability to generalize in identifying code vulnerabilities by discovering both n-day and zero-day vulnerabilities. To detect n-day vulnerabilities in the wild, we implement a vulnerability localization workflow on six C/C++ based IoT operating system repositories from GitHub. We begin by using JOERN to scan the repositories and extract functions.
As shown in Table \ref{tab:iot_data}, we detected a total of 24 vulnerabilities across five of the six repositories using T5-RCGCN, with no n-day vulnerabilities found in TinyOS. Comparing these results against other recent code vulnerability detection systems (Devign, ReVeAL), our proposed system detected significantly more vulnerabilities. The results also show that CWE-200 and CWE-120 vulnerabilities are present across multiple repositories, providing insight into the types of vulnerabilities prevalent in IoT devices.
In addition to n-day vulnerabilities, we discovered three zero-day vulnerable code samples in the IoT device operating systems. Our model, trained on a combined dataset of D2A and BigVul, successfully classified these samples and identified the vulnerable lines of code. The results were verified by our internal team of security experts. Of the three confirmed zero-day vulnerabilities, two were found in Zephyr and one in FreeRTOS. All three were categorized as CWE-119.

\subsection{Vulnerability Root Cause Evaluations}

We performed a second set of experiments to assess junior developers' ability to write secure code, both with and without tool assistance. These experiments followed the same methodology as our initial motivation experiments described in Section \ref{3_Motivation}. We divided the junior developers into two distinct, non-overlapping groups:

\noindent \textbf{Baseline Group}: This group, from our motivation experiments, used existing source code vulnerability analysis tools.

\noindent \textbf{T5-RCGCN Group}: This new group, consisting of different junior developers than those in the Baseline Group, used our proposed source code vulnerability analysis tool, T5-RCGCN.

The key difference between these groups lies in the capabilities of the tools they used:
\begin{itemize}
\item \textbf{Baseline Group tools:} Provided only Code Vulnerability Classification and Code Vulnerability Localization, as determined by existing source code vulnerability analysis tools.
\item \textbf{Our T5-RCGCN tool:} Provided Code Vulnerability Classification, Code Vulnerability Localization, \textit{and} Code Vulnerability Root Cause analysis, as determined by our T5-RCGCN tool.
\end{itemize}

By comparing these two groups, we aimed to evaluate the effectiveness of our proposed tool, particularly its additional root cause analysis feature, in improving junior developers' secure coding practices. The participants in the T5-RCGCN Group were entirely separate from those in the Baseline Group, ensuring no overlap between the two sets of junior developers. Additional information on the participant demographics for both groups can be found in Appendix Table \ref{tab:demo_moti}. We put all the survey questions in the footnote link. \footnote{https://docs.google.com/forms/d/e/1FAIpQLSdbQrbq3mIfIxqj6-GuJy7wqoXTTkmIQsyrz52JKLK4w53FqQ/viewform}

\subsubsection{Fixing Vulnerable Code With T5-RCGCN}

\begin{figure}[t]
        \centering
        
        \includegraphics[width=0.48\textwidth]{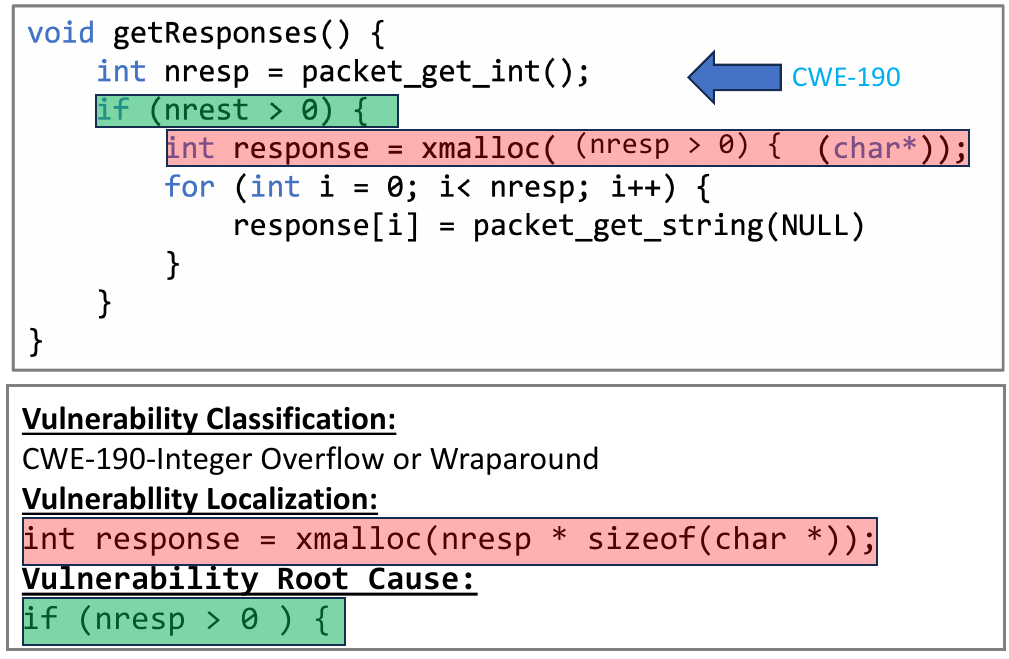}
        
    
    \caption{Sample source code provided to the participants depicted at the top and output at the bottom provided by T5-RCGCN. Our system takes in vulnerable code and outputs the vulnerability classification, vulnerability localization, and unlike other methods, our system also outputs the root cause of the vulnerability.}
    \label{fig:survey_our}
\end{figure}

Building upon our motivation experiments described in Section \ref{3_Motivation}, we conducted a new experiment to evaluate the effectiveness of our proposed T5-RCGCN tool for fixing vulnerable code. Participants in the T5-RCGCN Group were tasked with repairing the same five code functions as in Section \ref{sec:motivation_existing_tool_effectiveness}, which contain security vulnerabilities, mirroring the task given to the Baseline Group. We measured two key metrics: the correctness of the repairs and the time taken to complete them. Figure \ref{fig:survey_our} illustrates what participants in the T5-RCGCN Group saw: an example of vulnerable code and the output of T5-RCGCN, which provides Code Vulnerability Classification, Code Vulnerability Localization, and Code Vulnerability Root Cause analysis. This comprehensive output was designed to assist participants in understanding and addressing the vulnerabilities more effectively. Our results demonstrate a notable improvement for the T5-RCGCN Group compared to the previously established Baseline Group performance. On average, the T5-RCGCN Group successfully repaired 85.7\% of vulnerabilities, compared to the 56.7\% of successfully repaired vulnerabilities by those in the baseline group. We also find that this difference in repairs is statistically significant ($p < 0.01$) These findings suggest that the additional root cause analysis feature provided by our T5-RCGCN tool significantly enhances junior developers' ability to identify and fix security vulnerabilities more accurately and efficiently. ATable \ref{tab:qual_com_1} presents a detailed comparison of results between the T5-RCGCN Group and the Baseline Group for each code function, as well as the statistical significance of the differences observed. In addition to these results, we found that of those who successfully fixed the security vulnerabilities, the ones who used our tool were able to fix the vulnerability faster on average.

\begin{table}[b]
\centering
\caption{The average success rate of junior developers fixing five vulnerable code functions, and the average time taken to fix these code functions, measured in minutes. (** indicates $p < 0.01$).}
\resizebox{\columnwidth}{!}{
\begin{tabular}{l|l|l|l|l}
\hline
\textbf{Function name}    & \textbf{CWE Number} & \textbf{Baseline} & \textbf{T5-RCGCN} & \textbf{t-statistic} \\ \hline
getValueFromList & 125        & 56\%   & 89\%      &             \\ \cline{1-4}
*callHelper      & 416        & 50\%      & 89\%      &             \\ \cline{1-4}
SQLConnect       & 264        & 67\%      & 89\%      & 2.949**       \\ \cline{1-4}
readFile         & 416        & 78\%      & 100\%     &             \\ \cline{1-4}
*createBoard     & 20         & 33\%      & 62\%      &             \\ \hline
\end{tabular}
}
\label{tab:qual_com_1}
\end{table}

\subsubsection{Junior Developer Education}

\begin{figure}[t]
        \centering
        
        \includegraphics[width=0.5\textwidth]{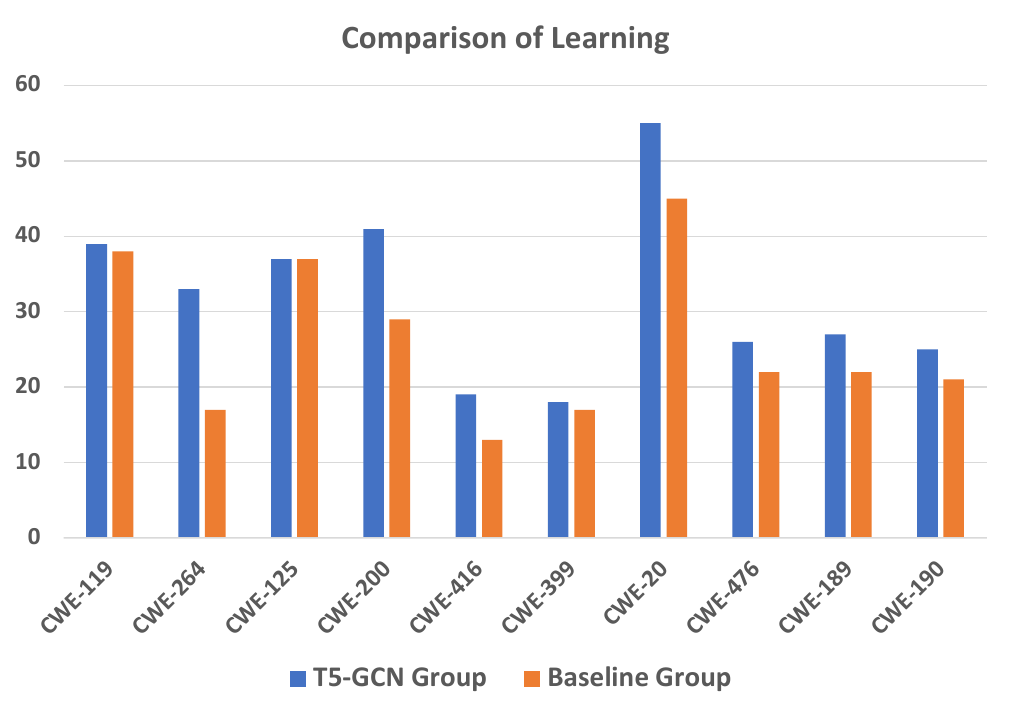}
        
    
    \caption{Performance of Developer Learning when Comparing T5-RCGCN vs. Baseline Group}
    \label{fig:learning_1}
\end{figure}

To assess the educational impact of source code vulnerability analysis tools, we conducted a follow-up experiment with the junior developers. This experiment revisited the ten function stubs initially presented in Section \ref{sec:without_tools}, where participants were originally tasked with solving and securing these stubs without any tool assistance. After using their respective tools (T5-RCGCN or baseline) to secure five different code samples, the developers were asked to reattempt the original ten function stubs, still without tool support. We measured the improvement in their performance as an indicator of learning effectiveness.
Figure \ref{fig:learning_1} presents the results of this experiment. The values shown represent the increase in the proportion of participants who correctly solved each function stub, calculated by subtracting the initial success rate from the post-tool-exposure success rate. Across all ten function stubs, the T5-RCGCN and Baseline groups demonstrated improvement, as evidenced by positive values.
Notably, the T5-RCGCN group consistently outperformed the Baseline group regarding improvement for each code task. This suggests that the T5-RCGCN source code vulnerability analysis tool may be more effective in educating junior developers about secure coding practices. We hypothesize that this enhanced learning is attributable to the Vulnerability Root Cause feature provided by our system, which helps teach junior developers how to identify better and understand the specific elements that render code vulnerable, thereby improving their ability to implement fixes.


The improvement of the developer performance for both the T5-RCGCN Group and the Baseline Group was most substantial for the CWE-20 (Improper Input Validation) vulnerability. The participants showed greater improvement on securing vulnerabilities like CWE-20 after having used tools previously compared to other vulnerabilities such as CWE-399 (Resource Management Errors), CWE-190 (Integer Overflow or Wraparound), and CWE-416 (Use After Free) due to several key factors. CWE-20 vulnerabilities typically involve straightforward and frequent patterns, such as missing or improper validation checks for user inputs. These patterns are easier to detect and fix because they follow a predictable structure that can be identified across various codebases. For example, replacing a direct assignment `strcpy(buffer, input)` with a safer alternative $strncpy(buffer, input, sizeof(buffer) - 1)$ is a clear, repetitive fix where the use of tools in the previous part in the study may have provided enough information to help the developers learn how to resolve the vulnerability. In contrast, CWE-399 issues are more complex as they involve intricate and context-specific resource management practices, like ensuring proper closure of file handles or preventing memory leaks, which require a deeper understanding of the resource lifecycle and context within the code.

CWE-119 (Improper Restriction of Operations within the Bounds of a Memory Buffer) also shows significant learning improvements across both the T5-RCGCN Group and the Baseline Group. Buffer overflow vulnerabilities are prevalent, especially in languages like C/C++, and involve clear, identifiable patterns. Fixes often include replacing unsafe functions with their safer counterparts or adding boundary checks, such as changing $`strcpy(buffer, src)`$ to $`strncpy(buffer, src, sizeof(buffer) - 1)`$. These straightforward and repetitive fixes are easier for the tools to identify and for the developers to learn from what these tools had identified. On the other hand, CWE-190 and CWE-416 involve subtler and more complex issues like ensuring integer operations do not exceed variable limits or managing the lifecycle of memory pointers to avoid use-after-free errors. These require the model to understand and analyze the broader code context and temporal aspects of operations, making them more challenging to address with a generic learning approach than the more frequent and structurally clear CWE-20 and CWE-119 vulnerabilities.

For CWE-264 and CWE-20, we see that the T5-RCGCN Group performed significantly better than the Baseline Group. These two vulnerabilities are unique in a way that they don't cause a program to crash or produce runtime errors directly when the code is vulnerable. Rather, CWE-264 occurs when a user gives unnecessary privileges to a file, which an attacker can use to execute malicious code. On the other hand, CWE-20 occurs when sensitive information like an incorrect password or username is provided to the user. This information helps an attacker guess whether they put the correct username, thereby shortening the search space. However, with the root cause information provided to the T5-RCGCN group, they can understand the problem lies with permission associated with a file or incorrect message provided. Therefore, they can solve the problem more efficiently than the baseline group.

\subsubsection{Other Observations}
From the follow-up questions after the coding exercises, we found that 13\% of the participants use ChatGPT extensively for code repair in our study. By analyzing the written code from these 13\% of the participants, we found that 18\% of the code generated by these participants had vulnerabilities. While this is lower than those who did not use ChatGPT, this is still a significant amount of vulnerabilities that were generated by on of the top performing LLMs. Another finding from the follow-up questions is that 42.3\% of the participants in the T5-RCGCN group responded that they feel more comfortable with analyzing and repairing source code vulnerability, compared to 34\% from the Baseline group after using these source code vulnerability analysis tools.


\section{Limitations and Future Work}
\label{8_limitations}


\paragraph{\textbf{Limited Scope of Code Vulnerability Samples}} In our study, we made concerted efforts to simulate a realistic scenario for "Grocery Store Management System" code development. However, this is a straightforward implementation of the system and may not completely capture the intricacies a developer would face in a real-world development condition. Furthermore, project size, team-based development, and quality assurance may not be fully captured in our survey design.

\paragraph{\textbf{Limited Participants}} Given the small sample size and apparent small effect sizes, establishing the statistical significance of the observed differences is challenging. We understand the limitations in terms of the small number of participants, but despite that why the results could still be considered significant since this group represents our target group who are early developers or recent graduates with minimal knowledge on code security.


In our future work, we plan to address these concerns by adding a broader group of participants, including professional developers, to analyze the effectiveness of our system. Moreover, to generalize the capabilities on root cause analysis, we plan to add multiple languages and see how our system performs.


\section{Conclusion}
\label{conclusion}
This study addresses the critical challenge of empowering junior developers to identify and fix security vulnerabilities in their code. Our comprehensive evaluation of existing software security tools revealed significant limitations, with developers able to secure only 36.2\% of vulnerable code across five vulnerability types. This finding underscores the pressing need for more effective solutions in the field of software security.
In response to these challenges, we developed T5-RCGCN, a source code vulnerability analysis tool that leverages the power of T5 language model embeddings and graph convolutional networks for improved vulnerability classification and localization. The integration of DeepLiftSHAP to pinpoint root causes of vulnerabilities addresses a crucial gap identified in our initial study, where developers struggled to identify the specific code segments responsible for security issues.
Our extensive evaluation, involving 56 junior developers and three source code vulnerability datasets, demonstrated the efficacy of T5-RCGCN. The tool improved developers' ability to proactively secure code by 14.6\% compared to previous methods, a significant advancement in the field. Moreover, the observed educational benefits of T5-RCGCN are particularly promising, as developers showed enhanced capability in securing code independently after using the tool.
These findings have important implications for both the immediate improvement of code security and the long-term enhancement of developer skills.

\bibliographystyle{unsrt}
\bibliography{main}

\begin{appendices}

\subsection{CWE Explanation}
\label{cwe_explain}
The following table provides a summary of the ten Common Weakness Enumerations (CWEs) addressed in our study. Each CWE represents a specific type of software weakness that could potentially lead to security vulnerabilities. Table \ref{tab:cwe_explain} provides all the CWE numbers with a small description we gathered from CWE \cite{cwe}.

\subsection{Survey Questionnaire}

To gain insights into our survey instruments and understand their implications, we put the survey questionnaire accessible through the provided link \url{https://anonymous.4open.science/r/Threat_Detection_Modeling-BB7B/Paper_Survey_A.csv}. However, for the sake of the anonymity of the participants, we removed their information, including their names, email addresses, and affiliations.


\subsection{Study Workflow}
\label{sec:study_workflow}

To assess the efficacy of our proposed system, we conducted a comprehensive evaluation involving 56 participants divided into control and assisted groups encompassing diverse educational and professional backgrounds with programming capabilities. The participants engaged in an online survey, wherein the survey's objective was presented as an evaluation of code writing quality. Initially, participants were questioned about their software development proficiency, including the extent of their coding experience and primary programming language expertise. Table \ref{tab:demo_moti} shows detailed information on the participants of our survey.

\begin{table}[t]
\centering
\caption{Demography of the Participants from the Assisted and Controlled Group}
\begin{tabular}{lrrr}
\hline
\multicolumn{1}{l|}{}                      & \multicolumn{1}{l|}{\textbf{Baseline Group}} & \multicolumn{1}{l|}{\textbf{T5-RCGCN Group}} & \multicolumn{1}{c}{\textbf{Total}} \\ \hline
\multicolumn{4}{c}{\textbf{Educational Background}}                                                                                                              \\ \hline
\multicolumn{1}{l|}{Junior (Year 1 and 2)} & \multicolumn{1}{r|}{5}                & \multicolumn{1}{r|}{6}                 & 11                                 \\ \hline
\multicolumn{1}{l|}{Senior (Year 3 and 4)} & \multicolumn{1}{r|}{6}                & \multicolumn{1}{r|}{8}                 & 14                                 \\ \hline
\multicolumn{1}{l|}{MS}                    & \multicolumn{1}{r|}{6}                & \multicolumn{1}{r|}{6}                 & 12                                 \\ \hline
\multicolumn{1}{l|}{PhD}                   & \multicolumn{1}{r|}{8}                & \multicolumn{1}{r|}{11}                & 19                                 \\ \hline
\multicolumn{4}{c}{\textbf{Years of Programming Experience}}                                                                                                     \\ \hline
\multicolumn{1}{l|}{0-2 Years}             & \multicolumn{1}{r|}{4}                & \multicolumn{1}{r|}{7}                 & 11                                 \\ \hline
\multicolumn{1}{l|}{2-5 Years}             & \multicolumn{1}{r|}{9}                & \multicolumn{1}{r|}{13}                & 22                                 \\ \hline
\multicolumn{1}{l|}{More than 5 Years}     & \multicolumn{1}{r|}{8}                & \multicolumn{1}{r|}{9}                 & 17                                 \\ \hline
\multicolumn{1}{l|}{Did not report}        & \multicolumn{1}{r|}{2}                & \multicolumn{1}{r|}{4}                 & 6                                  \\ \hline
\multicolumn{4}{c}{\textbf{Security Courses Taken}}                                                                                                              \\ \hline
\multicolumn{1}{l|}{Yes}                   & \multicolumn{1}{r|}{1}                & \multicolumn{1}{r|}{2}                 & 3                                  \\ \hline
\multicolumn{1}{l|}{No}                    & \multicolumn{1}{r|}{21}               & \multicolumn{1}{r|}{32}                & 53                                 \\ \hline
\multicolumn{1}{l|}{Total}                 & \multicolumn{1}{r|}{22}               & \multicolumn{1}{r|}{34}                & $\leftarrow$ N = 56                \\ \hline
\end{tabular}

\label{tab:demo_moti}
\end{table}

We initially directed participants to complete ten functions within an incomplete C source code file. These functions were intentionally crafted in a manner where potential vulnerabilities could be introduced if not approached with caution. Eligibility criteria for participation necessitated a fundamental understanding of the C programming language, ensuring that the tasks were within the participants' grasp. It is important to emphasize that the ten functions exclusively revolved around elementary C programming concepts and excluded intricate topics like data structures. In cases where participants did not fulfill any of the functions, they were gracefully guided to the survey's conclusion. This approach aimed to maintain a consistent and relevant experience for all participants. Listing \ref{lst:lst2} shows the header file and the structure we provided to the participants.

\begin{table*}[t]
\centering
\caption{A short description on the categories of CWEs we Analyzed}
\begin{tabular}{@{}l|l@{}}
\toprule
\textbf{CWE Number} & \textbf{Description}                                                    \\ \midrule
CWE-119             & Improper Restriction of Operations within the Bounds of a Memory Buffer \\
CWE-264             & Permissions, Privileges, and Access Controls                            \\
CWE-125             & Out-of-bounds Read                                                      \\
CWE-200:            & Exposure of Sensitive Information to an Unauthorized Actor              \\
CWE-416             & Use After Free                                                          \\
CWE-399             & Resource Management Errors                                              \\
CWE-20              & Improper Input Validation                                               \\
CWE-476             & NULL Pointer Dereference                                                \\
CWE-189             & Numeric Errors                                                          \\
CWE-190             & Integer Overflow or Wraparound                                          \\ \bottomrule
\end{tabular}
\label{tab:cwe_explain}
\end{table*}

\begin{lstlisting}[style=CStyle, label={lst:lst2}, caption=The included headers\, defined constants\, and defined structures used in the survey.]
#include <stdlib.h>
#include <stdio.h>
#include <string.h>

typedef int bool;

#define TRUE 0
#define FALSE 1

/**
 * User of the grocery store system.
 */
typedef struct
{
    char *username; // Username of the user.
    char *password; // Password of the user.
} User;
\end{lstlisting}

Furthermore, \ref{lst:lst3} and \ref{lst:lst4} exemplify some other code samples provided to the participants. These are empty functions with a proper description of the problem, a sample test case of the solution, and a clarification on the return type. Inside the function body, there is an option to put the starting and ending times to measure how long it took them to complete the function.

If participants made partial progress on the coding assignment, a series of inquiries ensued regarding their familiarity with code vulnerabilities and the extent of their formal cybersecurity training. Following the collection of security-related background information, participants were assigned to either "Form A" or "Form B," with "Form A" representing the assisted group and "Form B" serving as the control group. The survey structure was thoughtfully arranged such that every alternate participant received "Form B," a measure intended to achieve an equitable distribution of approximately 50\% between the control and assisted groups.

\begin{lstlisting}[style=CStyle, label={lst:lst3}, caption=A function to write the updated price of each items to a file]
void exportPrices(char **itemStrings, int numItems, char *filename)
{
    /**
     * Problem:
     *  Export the contents of the grocery store system to a file. Each line of the file should
     * contain one grocery store item.
     *
     * Example:
     *  exportPrices({"Apple - $1.25", "Orange - $0.75"}, 2, "out.txt") the contents of out.txt would
     * look like:
     * Apple - $1.25
     * Orange - $0.75
     *
     * Parameters:
     *  itemStrings: an array of grocery store items formatted as "Item - $X.XX".
     *  numItems: the length of itemStrings.
     *  filename: the name of the output file to export prices to.
     */
    //  --> PARTICIPANT: ENTER HERE TIME STARTED <--
}
//  --> PARTICIPANT: ENTER HERE TIME COMPLETED <--
\end{lstlisting}

Irrespective of the assigned form, participants were presented with five distinct C functions, which they were asked to repair. The control group only had visibility to the CWE class of the vulnerability and the vulnerable line, as depicted in Figure \ref{fig:survey_sota}. However, the assisted group was provided with the CWE category of the vulnerability with a static description, the vulnerable line, and the root cause of the vulnerability as depicted in Figure \ref{fig:survey_our}.

Upon completing the vulnerability assessment task, participants were further probed with questions assessing their confidence levels in mitigating code vulnerabilities and whether they sought assistance from ChatGPT for code composition. The survey concluded with the collection of pertinent demographic data from each participant.

\begin{lstlisting}[style=CStyle, label={lst:lst4}, caption=A function that primarily checks if the participant scanned values from a string correctly.]
double extractPrice(char *itemString)
{
    /**
     * Problem:
     *  In the grocery store, prices are stored in the system in the format of "Item - $X.XX".
     * This function should extract the price from the string and return it as a double.
     *
     * Example:
     *  extractPrice("Apple - $1.25") returns 1.25
     *
     * Parameters:
     *  itemString: a grocery store item formatted as "Item - $X.XX".
     *
     * Returns:
     *  the price of the item as a double.
     */
    //  --> PARTICIPANT: ENTER HERE TIME STARTED <--
    return 0.0;
}
//  --> PARTICIPANT: ENTER HERE TIME COMPLETED <--
\end{lstlisting}

    \label{9_appendix}
\end{appendices}



%



\end{document}